\documentclass[a4paper,11pt]{article}
\usepackage{pos}
\usepackage{amsmath}
\usepackage{epsfig}
\usepackage{slashed}
\usepackage{xcolor}
\usepackage{graphicx}
\usepackage{array}
\usepackage{epic}
\usepackage{eepic}
\usepackage{epsfig}
\usepackage{latexsym}
\usepackage[export]{adjustbox}
\usepackage{float}
\usepackage{multirow}
\usepackage{hyperref}
\usepackage[caption=false]{subfig}

\hypersetup{colorlinks=true,citecolor=red,linkcolor=blue,urlcolor=blue}

\title{Precise LHC limits on the $\rm{U}_1$ leptoquark parameter space}
\author*[a]{Arvind Bhaskar}
\author[a]{Diganta Das}
\author[b]{Tanumoy Mandal}
\author[a]{Subhadip Mitra}
\author[a]{Cyrin Neeraj}

\affiliation[a]{Center for Computational Natural Sciences and Bioinformatics, International Institute of Information Technology, Hyderabad 500 032, India}

\affiliation[b]{Indian Institute of Science Education and Research Thiruvananthapuram, Vithura, Kerala, 695 551, India}

\emailAdd{arvind.bhaskar@research.iiit.ac.in}
\emailAdd{diganta.das@iiit.ac.in}
\emailAdd{tanumoy@iisertvm.ac.in}
\emailAdd{subhadip.mitra@iiit.ac.in}
\emailAdd{cyrin.neeraj@research.iiit.ac.in}

\abstract{A TeV scale leptoquark (LQ) is one of the promising explanations of the recent anomalies in the semileptonic decays of $B$ mesons. Among the various LQs, the vector $\rm{U}_1$ is capable of explaining the anomalies in both $R_{D^{(*)}}$ and $R_{K^{(*)}}$ observables. We use the current LHC data to put bounds on the parameter space of $\rm{U}_1$ relevant for the anomalies. Precise bounds are drawn by recasting the latest $\tau\tau$ and $\mu\mu$ searches by the ATLAS and CMS collaborations. We find that it is imperative to include the resonant production modes for obtaining limits in the low mass regions. For higher mass points, the non-resonant production modes play a dominant role. }

\FullConference{%
  *** The European Physical Society Conference on High Energy Physics (EPS-HEP2021), ***\\
  *** 26-30 July 2021 ***\\
  *** Online conference, jointly organized by Universität Hamburg and the research center DESY ***
}

\begin{document}
\maketitle

%%%%%%%%%%%%%%%%%%%%%%%%%%%%%%%%%%%%%%%%%%%%%%%%%%%%%%%%
%%%%%%%%%%%%%%%%%%%%%%%%%%%%%%%%%%%%%%%%%%%%%%%%%%%%%%%%
\section{Introduction}
%%%%%%%%%%%%%%%%%%%%%%%%%%%%%%%%%%%%%%%%%%%%%%%%%%%%%%%%
%%%%%%%%%%%%%%%%%%%%%%%%%%%%%%%%%%%%%%%%%%%%%%%%%%%%%%%%

The discrepancies among the theoretical and experimental values of $R_{D^{(*)}}$ and $R_{K^{(*)}}$ observables might lead to a pathway to the discovery of beyond the standard model (BSM) physics. They are defined as follows, 
\begin{align}
R_{D^{(*)}}  =  \dfrac{{\mathcal {B}}(B\rightarrow D^{(*)}\tau\bar\nu)}{{\mathcal {B}}(B\rightarrow D^{(*)}\hat{\ell}\bar\nu)} \quad \mbox{and}\quad  R_{K^{(*)}} = \dfrac{{\mathcal {B}}(B\rightarrow K^{(*)}\mu^+\mu^-)}{{\mathcal {B}}(B\rightarrow K^{(*)}e^+e^-)}\ .\label{eq:anomalies}
\end{align}
The recent  measurement of  $R_{K^{(*)}}$ are smaller than the theoretical prediction by $3.1\sigma$ \cite{Bordone:2016gaq, Hiller:2003js, Blanke:2018sro, Calibbi:2015kma, Calibbi:2017qbu, Crivellin:2018yvo}. Whereas the measurements of the $R_{D^{(*)}}$ observables show a combined excess of $3.1\sigma$ \cite{HFLAV:2019otj} than the Standard Model (SM) value.
From the many models proposed to explain the  anomalies, we focus on the $U_1\equiv(\mathbf{3},\mathbf{1},2/3)$ leptoquark (LQ) model that can  explain both the anomalies simultaneously. LQs are colored triplet, scalar or vector bosons. They carry both lepton and baryon numbers and decay to the leptons and quarks. LQs appear in many models like, $SU(4)$ Pati-Salam models, Technicolor models, RPV SUSY models, etc. Their coupling to SM leptons and quarks makes them an ideal candidate to explain the $B$-anomalies. %In this work we have considered the vector LQ (vLQ), $U_1\equiv(\mathbf{3},\mathbf{1},2/3)$. It can explain both $R_{D^{(*)}}$ and $R_{K^{(*)}}$ anomalies simultaneously. 

%%%%%%%%%%%%%%%%%%%%%%%%%%%%%%%%%%%%%%%%%%%%%%%%%%%%%%%%%%%%%%%%%%%%%%%%%%%%%%%%%%%%%%%%%%%%%%%%%%%%%%%%%%%%%%%%%%%%%%%%%%%%%%%%
\section{The $\rm{U}_1$ Lagrangian}
%%%%%%%%%%%%%%%%%%%%%%%%%%%%%%%%%%%%%%%%%%%%%%%%%%%%%%%%%%%%%%%%
%%%%%%%%%%%%%%%%%%%%%%%%%%%%%%%%%%%%%%%%%%%%%%%%%%%%%%%%%%%%%%%%

The Lagrangian for the interactions between $\rm{U_1}$ and the SM quarks and leptons is given as, 
\begin{equation}
\label{eq:GenLagU1}
\mathcal{L} \supset
x^{LL}_{1~ij}~\bar{Q}^{i}\gamma_{\mu}U^{\mu}_{1}P_{L}L^{j} + x_{1~ij}^{RR}~\bar{d}^i_{R}\gamma_{\mu}U^{\mu}_{1}P_{R}\ell^j_{R} + \textrm{h.c.}
\end{equation}
Here, $Q_i$ and $L_j$ are the SM left-handed quark and lepton doublets, respectively. The $d_{R}^{i}$ and $\ell_{R}^{j}$ are the down-type right-handed quarks and leptons, respectively. The indices $i,j=\{1,\ 2,\ 3\}$ stand for the quark and lepton generations. Thus, the $x_{1~ij}^{LL}$ and $x_{1~ij}^{RR}$ are $3\times 3$ matrices in the flavour space. Here, we consider only real values of the couplings in the matrices and set all couplings but the ones involved in the $b\to c\tau\bar{\nu}$ (i.e., $\lambda^L_{23}$, $\lambda^{L/R}_{33}$) and $b\to s\mu^+\mu^-$ (i.e., $\lambda^{L/R}_{22}$, $\lambda^{L/R}_{32}$) decays to be zero.
\begin{equation}
x^{LL}_1 = 
\begin{pmatrix}
0 & 0 & 0 \\
0 & {\lambda^L_{22}} & \lambda^L_{23} \\
0 & {\lambda^L_{32}} & \lambda^L_{33}
\end{pmatrix};~~
x^{RR}_1 = 
\begin{pmatrix}
0 & 0 & 0 \\
0 & {\lambda^R_{32}} & 0 \\
0 & {\lambda^R_{32}} & \lambda^R_{33}
\end{pmatrix}. \label{eq:rkcouplings}
\end{equation}

%%%%%%%%%%%%%%%%%%%%%%%%%%%%%%%%%%%%%%%%%%%%%%%%%%%%%%%%%%%%%%%%%%%%%%%%%%%%%%%%%%%%%%%%%%%%%%%%%%%%%%%%%%%%%%%%%%%%%%%%%%%%%%
\subsection{The $R_{D^{(*)}}$ \& $R_{K^{(*)}}$ scenarios}
%%%%%%%%%%%%%%%%%%%%%%%%%%%%%%%%%%%%%%%%%%%%%%%%%%%%%%%%%%%%%%%
%%%%%%%%%%%%%%%%%%%%%%%%%%%%%%%%%%%%%%%%%%%%%%%%%%%%%%%%%%%%%%%

To explain the $R_{D^{(*)}}$ scenarios, the new physics needs to contribute to the $b\to c\tau\bar{\nu}$ transition. Hence, in our case, the $\rm{U}_1$ needs to couple with both $c$-$\nu$ and $b$-$\tau$ pairs. To obtain the required couplings, we first consider some minimal single-coupling scenarios and then move on to more complex multi-coupling scenarios. In the first single-coupling scenario, we
consider only $\lambda^L_{23}$ to be non-zero in the following manner,
\begin{equation}
\mathcal{L} \supset \lambda_{23}^{L}[\bar{c}_L\gamma_{\mu}\nu_{L} + \bar{s}_L \gamma_{\mu}\tau_{L})] U_{1}^{\mu}
= \lambda_{23}^{L}[\bar{c}_L\gamma_{\mu}\nu_{L} + (V^*_{cd}\bar{d}_L + V^*_{cs}\bar{s}_L + V^*_{cb}\bar{b}_L) \gamma_{\mu}\tau_{L})] U_{1}^{\mu},
\end{equation}
i.e., the $c$-$\nu$ coupling is generated directly but the $b$-$\tau$ coupling is an effective one, generated by the mixing among the down-type quarks. In ref.~\cite{Bhaskar:2021pml}, this scenario is called RD1A. 
Similarly, if we put all but $\lambda^L_{33}$ to be zero, i.e., we generate the $b$-$\tau$ coupling directly and $c$-$\nu$ coupling via quark mixing, we get the scenario RD1B with the following lagrangian,
\begin{equation}
\mathcal{L} \supset \lambda_{33}^{L}[\bar{t}_L \gamma_{\mu}\nu_{L} + \bar{b}_L\gamma_{\mu}\tau_{L}] U_{1}^{\mu}
= \lambda_{33}^{L}[(V_{ub}\bar{u}_L + V_{cb}\bar{c}_L + V_{tb}\bar{t}_L) \gamma_{\mu}\nu_{L}) + \bar{b}_L\gamma_{\mu}\tau_{L}] U_{1}^{\mu}.
\end{equation}
The possible single and multi-coupling scenarios have been summarized in Table~\ref{tab:RDobs}. 
There we also show some of the possible scenarios one can construct to accommodate the $R_{K^{(*)}}$ anomalies. These scenarios may seem similar from an effective-field-theory point of view, but their LHC signatures are quite different. Further details on these scenarios can be found in ref.~\cite{Bhaskar:2021pml}.

\begin{table}[t!]
\caption{Summary of the single and multi-coupling scenarios to explain the $R_{D^{(*)}}$ and $R_{K^{(*)}}$ anomalies (from ref.~\cite{Bhaskar:2021pml}).}
\begin{center}
{\linespread{1.3}\footnotesize
\begin{tabular*}{\textwidth}{l@{\extracolsep{\fill}}ccclcccc}
\hline
$R_{D^{(*)}}$ scenarios & $\lambda_{c\nu}^L$ & $\lambda_{b\tau}^L$ & $\lambda_{b\tau}^R$ &$R_{K^{(*)}}$ scenarios&$\lambda_{s\mu}^L$ & $\lambda_{b\mu}^L$ & $\lambda_{s\mu}^R$ & $\lambda_{b\mu}^R$ \\  

\hline\hline
RD1A  & $\lambda_{23}^L$ & $V_{cb}^*\lambda_{23}^L$ & $-$ & RK1A  & $V_{cs}^*\lambda_{22}^L$ & $V_{cb}^*\lambda_{22}^L$ & $-$ & $-$ \\
RD1B  & $V_{cb}\lambda_{33}^L$  & $\lambda_{33}^L$ & $-$ & RK1B  & $V_{ts}^*\lambda_{32}^L$ & $V_{tb}^*\lambda_{32}^L$ & $-$ & $-$ \\
  &   &    &   & RK1C  & $-$ & $-$ & $V_{cs}\lambda_{22}^R$ & $V_{cb}\lambda_{22}^R$ \\
&   &    &   & RK1D  & $-$ & $-$ & $V_{ts}\lambda_{32}^R$ & $V_{tb}\lambda_{32}^R$   \\
\hline
RD2A  & $V_{cs}\lambda_{23}^{L} + V_{cb}\lambda_{33}^{L}$ & $\lambda_{33}^{L}$ & $-$ &  RK2A  & $\lambda_{22}^L$ & $\lambda_{32}^L$ & $-$ & $-$  \\
RD2B  & $V_{cs}\lambda_{23}^{L}$  &  $-$  &  $\lambda_{33}^{R}$ & RK2B  & $\lambda_{22}^L$ & $-$ & $-$ & $\lambda_{32}^R$  \\
&   &    &   & RK2C & $-$ & $\lambda_{32}^L$ & $\lambda_{22}^R$ & $-$  \\
&   &   &   & RK2D  & $-$ & $-$ & $\lambda_{22}^{R}$ & $\lambda_{32}^{R}$ \\\hline
RD3  & $V_{cb}\lambda^L_{33} + V_{cs} \lambda^L_{23}$  &  $\lambda^L_{33}$  &  $\lambda_{33}^{R}$ & RK4  & $\lambda_{22}^{L} $ & $\lambda_{32}^{L}$ & $\lambda_{22}^{R}$ & $\lambda_{32}^{R}$ \\
\hline
\end{tabular*}
}
\label{tab:RDobs}
\end{center}
\end{table}

% \begin{table}[t!]
% \caption{Summary of single and multi-coupling scenarios explaining $R_{D^{(*)}}$}
% \begin{center}
% {\linespread{1.5}\footnotesize
% \begin{tabular*}{\textwidth}{l@{\extracolsep{\fill}}ccc}
% \hline
% $R_{D^{(*)}}$ scenarios & $\lambda_{c\nu}^L$ & $\lambda_{b\tau}^L$ & $\lambda_{b\tau}^R$ \\  

% \hline\hline
% \hyperlink{sce:rd1a}{RD1A}  & $\lambda_{23}^L$ & $V_{cb}^*\lambda_{23}^L$ & $-$ \\
% %
% \hyperlink{sce:rd1a}{RD1B}  & $V_{cb}\lambda_{33}^L$  & $\lambda_{33}^L$ & $-$ \\
% %

% \hline
% %
% \hyperlink{sce:rd2a}{RD2A}  & $V_{cs}\lambda_{23}^{L} + V_{cb}\lambda_{33}^{L}$ & $\lambda_{33}^{L}$ & $-$ \\
% %
% \hyperlink{sce:rd2b}{RD2B}  & $V_{cs}\lambda_{23}^{L}$  &  $-$  &  $\lambda_{33}^{R}$ \\
% %
% \hline
% %
% \end{tabular*}}
% \label{tab:RDobs}
% \end{center}
% \end{table}

%%%%%%%%%%%%%%%%%%%%%%%%%%%%%%%%%%%%%%%%%%%%%%%%%%%%%%%%%%%%%%%%%%%%%%%%%%%%%%%%%%%%%%%%%%%%%%%%%%%%%%%%%%%%%%%%%%%%%%%%%%%%%%
\section{Production at the LHC and data recast}
%%%%%%%%%%%%%%%%%%%%%%%%%%%%%%%%%%%%%%%%%%%%%%%%%%%%%%%%%%%%%%%
%%%%%%%%%%%%%%%%%%%%%%%%%%%%%%%%%%%%%%%%%%%%%%%%%%%%%%%%%%%%%%%

We generate Monte Carlo events for all LQ production modes in the $R_{D^{(*)}}$ and $R_{K^{(*)}}$ scenarios. While recasting the LHC dilepton data~\cite{ATLAS:2020zms,CMS:2021ctt} we  systematically combine the events that can contribute to these searches~\cite{Mandal:2012rx,Mandal:2015vfa,Mandal:2016csb,Mandal:2018kau,Aydemir:2019ynb}. The $U_1$ can be produced at the LHC via two types of processes---resonant productions (like pair and the single productions) and non-resonant productions (like $t$-channel LQ exchange). In the RD1A (RD1B) scenario, the  pair production mode, which is mostly QCD mediated, gives the following final states, 
\begin{equation}
\label{eq:pairRD1A}
pp\to U_1 U_1 \rightarrow s\tau \, s\tau (b\tau \, b\tau)\equiv \tau\tau + 2j.
\end{equation}
The relevant single-production channels in the $R_{D^{(*)}}$ ($R_{K^{(*)}}$) scenarios can be written as
\begin{equation}
\label{eq:singRD1A}
pp\to U_1\tau(\mu) + U_1\tau(\mu) j  \rightarrow \tau\tau j(\mu\mu j) + \tau\tau jj(\mu\mu jj) \equiv \tau\tau(\mu\mu) + n\,j.
\end{equation}
The single productions depend on the $U_1$-$\ell$-$q$ couplings ($\lambda$) and play a significant role in the limits.
The $\lambda$-dependent non-resonant process where a $\rm{U}_1$ is exchanged in the $t$-channel can lead to $\tau\tau$ ($\mu\mu$) final states. 
Moreover, this process interferes with the SM background processes like $pp\to Z\gamma^*\to\tau\tau (\mu\mu)$. 
%Thus we ensure that all production modes lead to $\tau\tau$ or $\mu\mu$  in the final states. 

Since there is destructive interference between the BSM process and the SM background, we use a $\chi^2$ test instead of a simple recast to recast the latest $\tau\tau$ ($\mu\mu$) search data from ATLAS~\cite{ATLAS:2020zms} (CMS~\cite{CMS:2021ctt}) to obtain bounds. The $\chi^2$ can be expressed as follows~\cite{Bhaskar:2021pml}, 
\begin{equation}
            \chi^2 = \sum_i^{\rm bins} \left(\frac{ \left[N_{p}+N^{incl}_s+N_t-N_{\times}\right]^i + N_{BG}^i - N_{D}^i }{\sqrt{\left(\Delta N^i_{Stat}\right)^2+\left(\Delta N^i_{Syst}\right)^2}}\right)^2 
            \label{eqnchisq}
\end{equation}
where $N_{p}$, $N^{incl}_s$, $N_t$, $N_{\times}$, $N_{BG}$, and $N_{D}$ are the number of events from pair production, single production, pure-BSM ($t$-channel $U_1$ exchange), the BSM-SM interference, the SM background, and the data, respectively and $N^i_{Stat}=\sqrt{N_D^i}$ and $N^i_{Syst}=0.1\, N_D^i$ are the statistical and systematic uncertainties. For a range of $U_1$ masses, we use the total transverse mass distribution from the $\tau\tau$ data and the invariant mass distribution from the  $\mu\mu$ data to find the best fit parameters (i.e., couplings) and the $2\sigma$ exclusion limits on them.

%%%%%%%%%%%%%%%%%%%%%%%%%%%%%%%%%%%%%%%%%%%%%%%%%%%%%%%%%%%%%%%%%%%%%%%%%%%%%%%%%%%%%%%%%%%%%%%%%%%%%%%%%%%%%%%%%%%%%%%%%%%%%%%%%%%%%%%%%%%%%%%%
\section{Results}
%%%%%%%%%%%%%%%%%%%%%%%%%%%%%%%%%%%%%%%%%%%%%%%%%%%%%%%%%%%%%%%%%%%%%%%%
%%%%%%%%%%%%%%%%%%%%%%%%%%%%%%%%%%%%%%%%%%%%%%%%%%%%%%%%%%%%%%%%%%%%%%%%

\begin{figure*}[!t]
\centering
\captionsetup[subfigure]{labelformat=empty}
\subfloat[\quad\quad\quad(a)]{\includegraphics[width=0.4\textwidth]{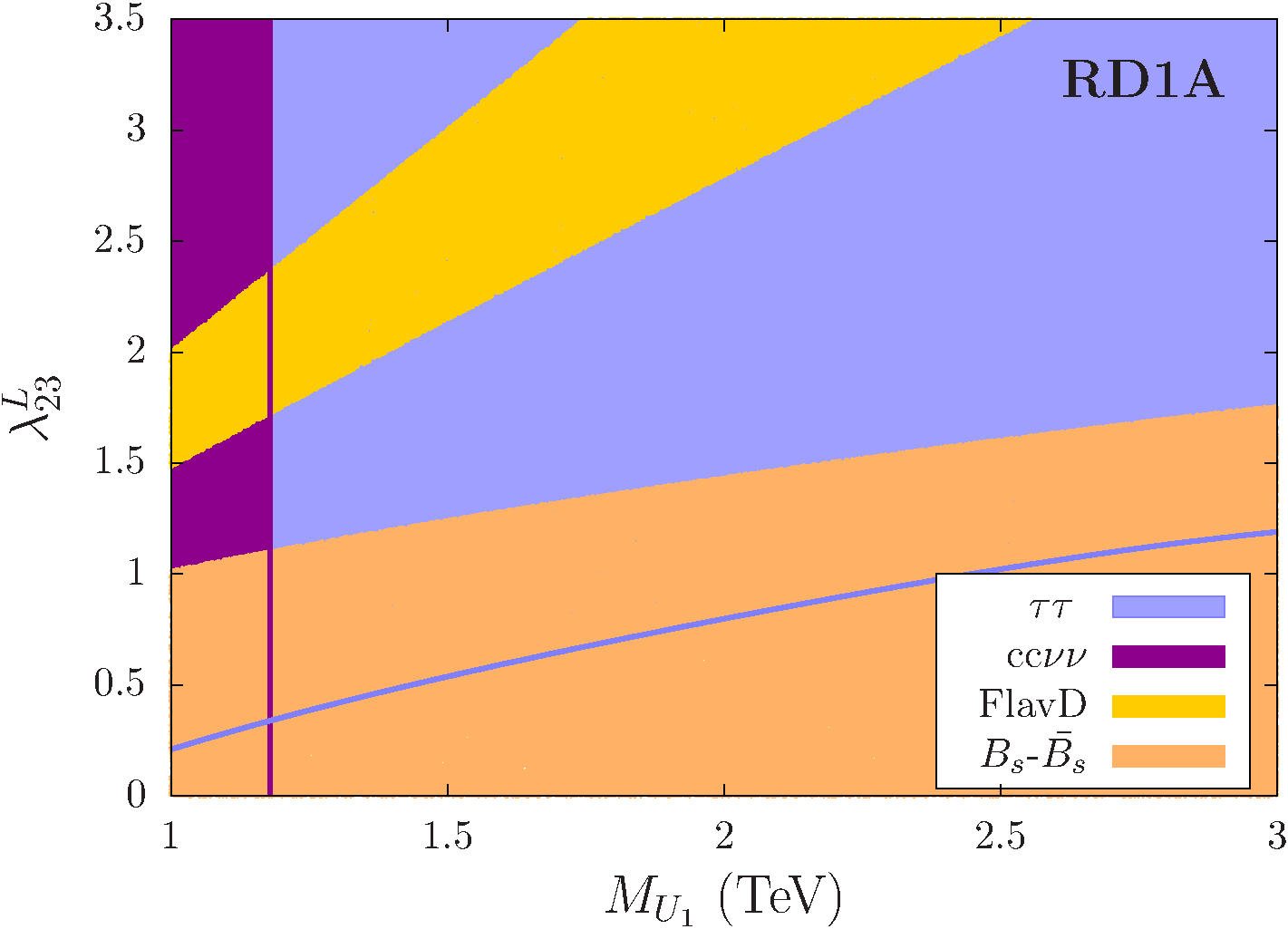}\label{fig:RD1alla}}\quad\quad
\subfloat[\quad\quad\quad(b)]{\includegraphics[width=0.4\textwidth]{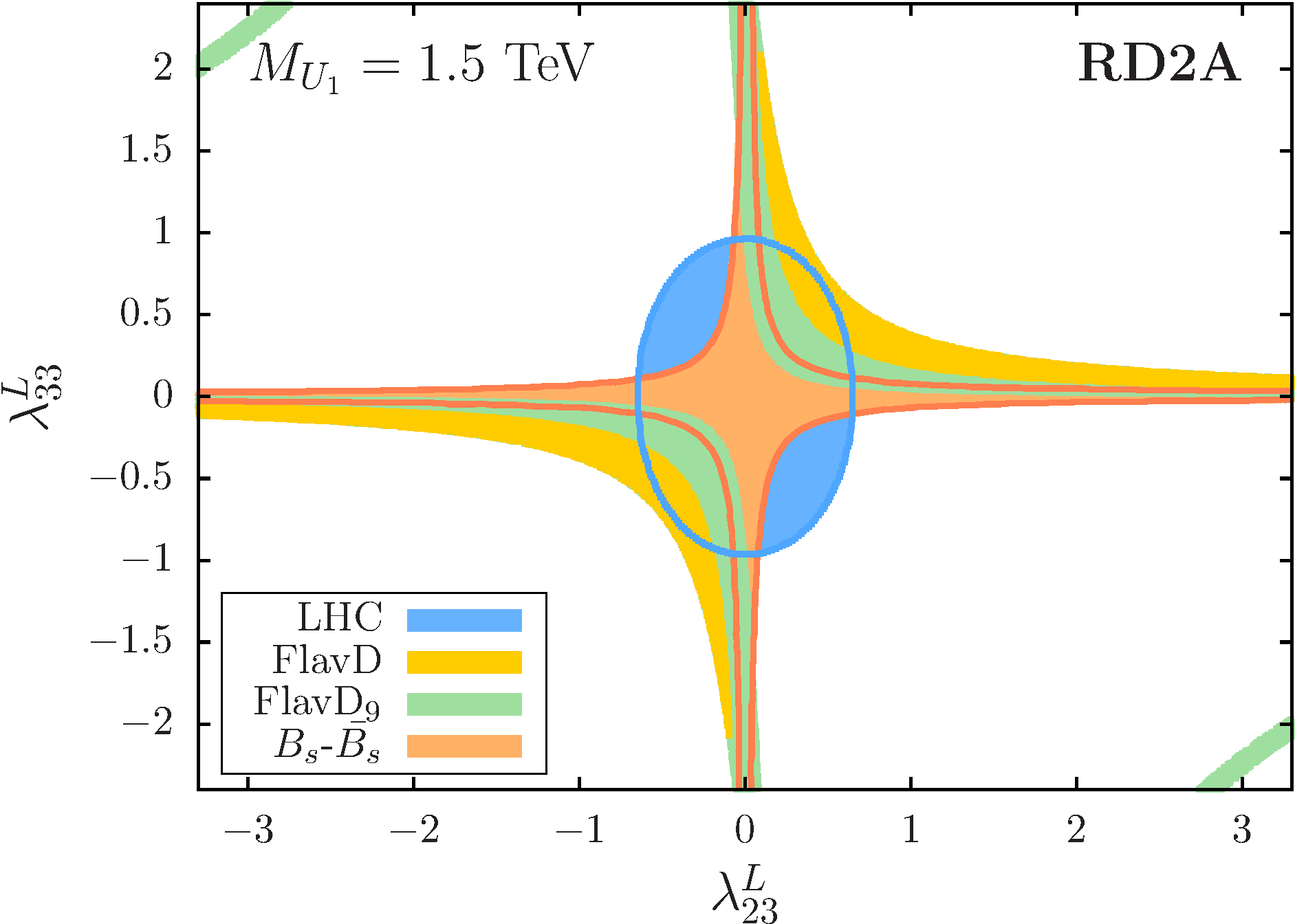}\label{fig:RD1allb}}
\caption{(a) The region (in purple) excluded by the LHC $\tau\tau$ data and the region (in yellow) favoured by the flavour observables in a $\lambda^L_{23}$-$M_{{U}_1}$ plane. (b) For a $1.5$TeV ${U}_1$, the blue color region is allowed by the LHC $\tau\tau$ data (from ref.~\cite{Bhaskar:2021pml}).}
\label{fig:RD1all}
\end{figure*}

\begin{figure*}[!t]
\centering
\captionsetup[subfigure]{labelformat=empty}
\subfloat[\quad\quad\quad(a)]{\includegraphics[width=0.4\textwidth]{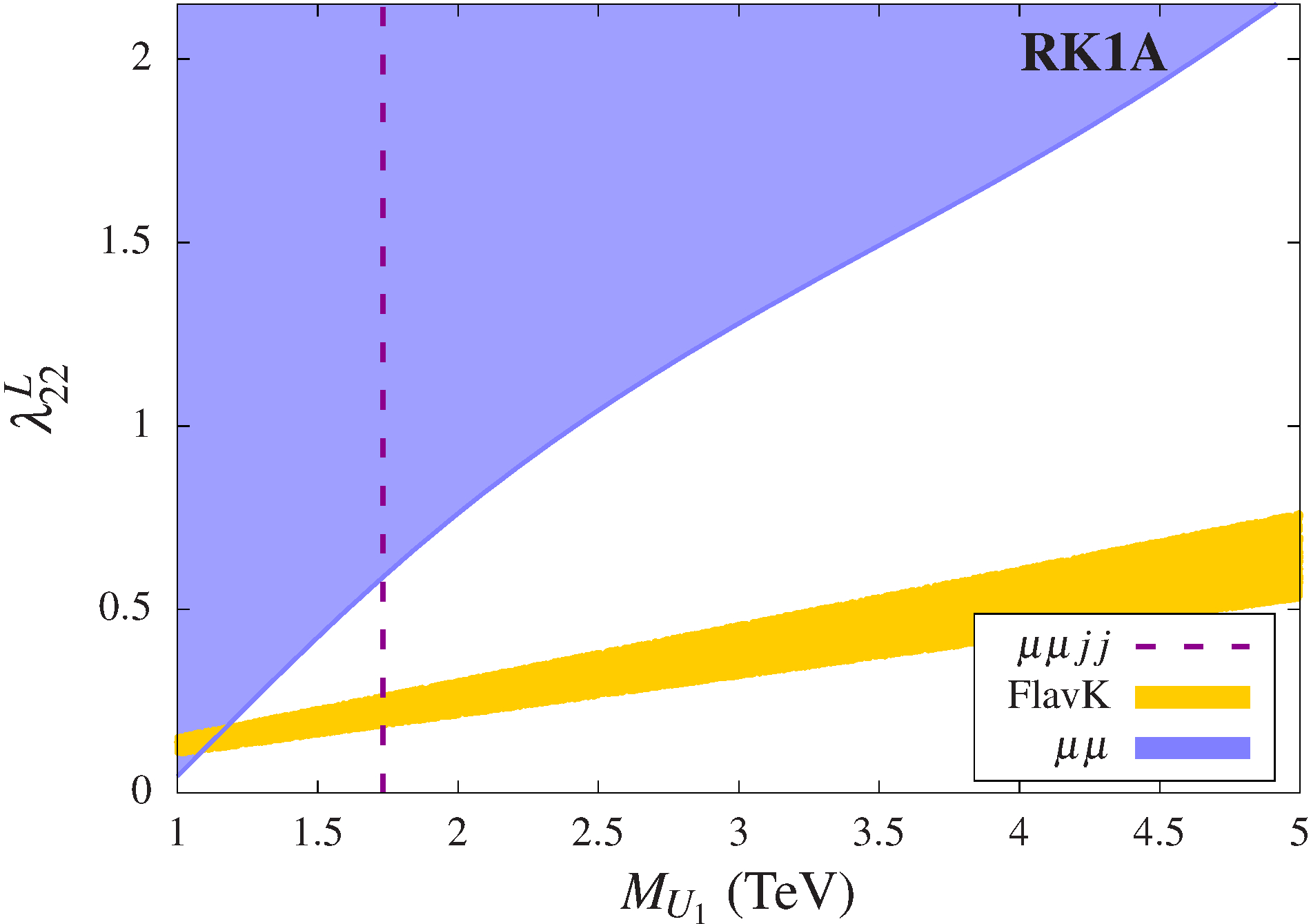}\label{fig:RK12alla}}\quad\quad
\subfloat[\quad\quad\quad(b)]{\includegraphics[width=0.4\textwidth]{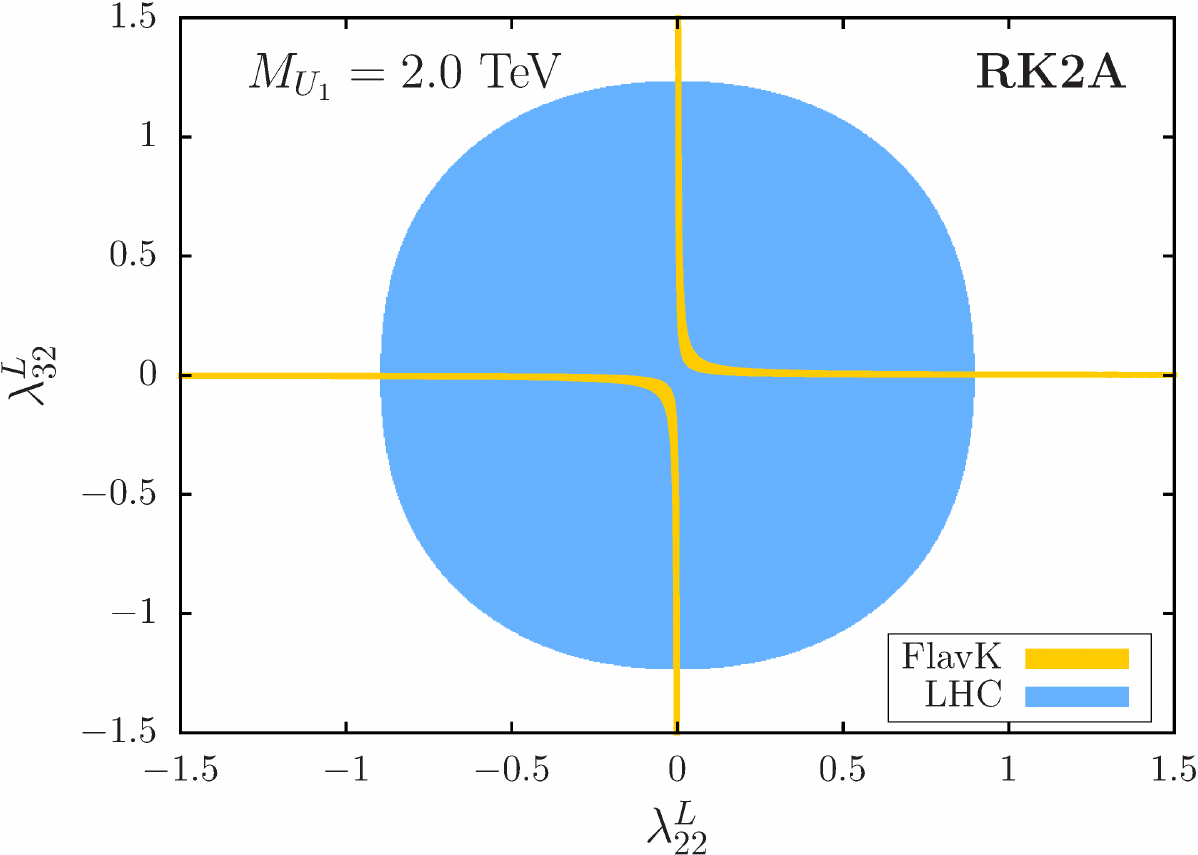}\label{fig:RK12allb}}
\caption{a) Excluded region (purple) by the LHC and the $R_{K^{(*)}}$-favoured region (yellow). (b) For a $2.0$ TeV $U_1$, the regions allowed by the LHC (in blue) and relevant to the $R_{K^{(*)}}$ observables (in yellow)~\cite{Bhaskar:2021pml}.}
\label{fig:RK12all}
\end{figure*}

\begin{figure*}[!t]
\captionsetup[subfigure]{labelformat=empty}
\centering\includegraphics[width=0.4\textwidth]{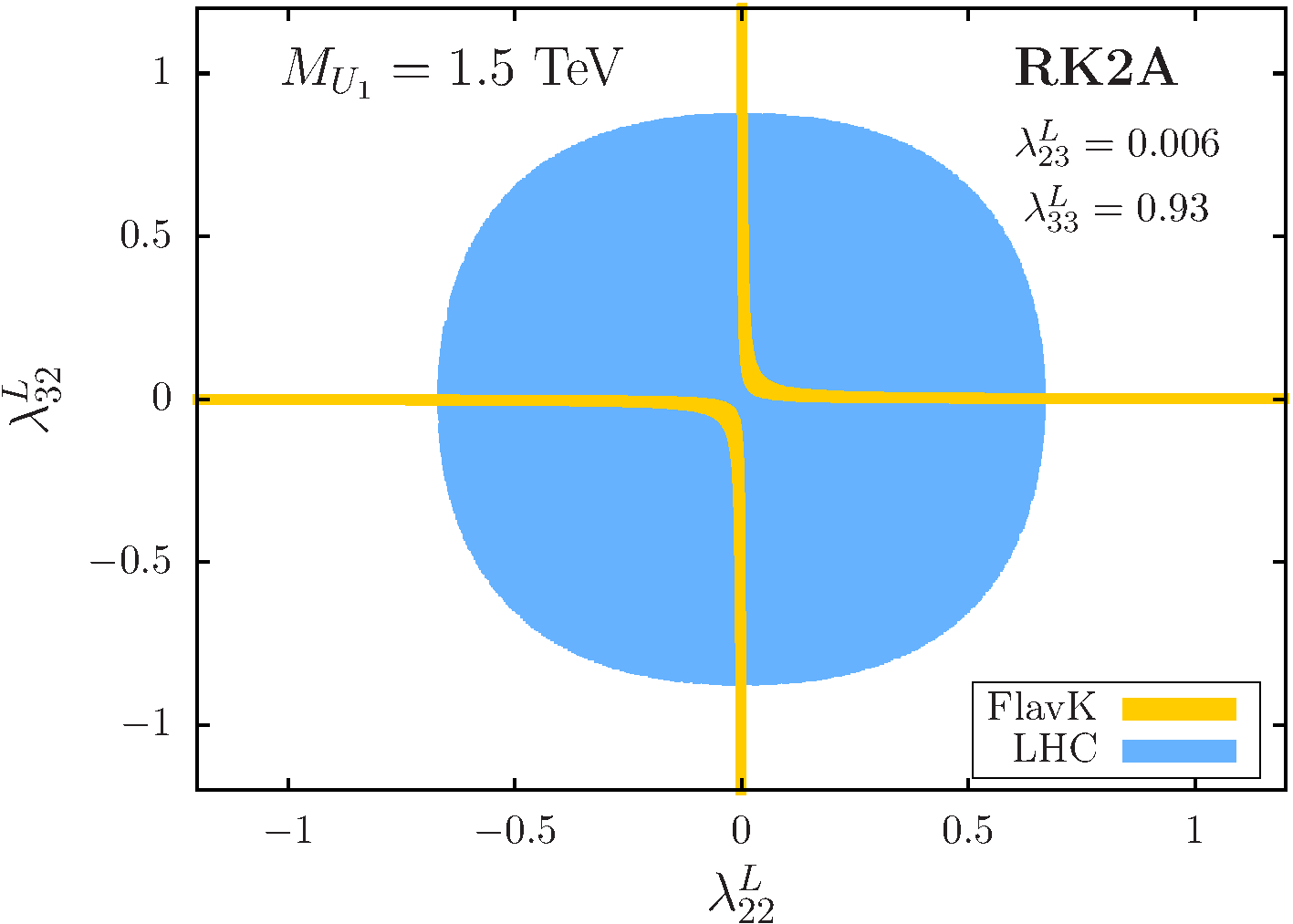}\label{fig:RK2Aa}
\caption{Region in the $\lambda^L_{22}$-$\lambda^L_{32}$ plane for a $1.5$ TeV $U_1$ allowed by the LHC data (blue) and simultaneously explaining the $R_{D^{(*)}}$ and $R_{K^{(*)}}$ anomalies (yellow)~\cite{Bhaskar:2021pml}.}
\label{fig:RK1500}
\end{figure*}

In Figure~\ref{fig:RD1alla}, we show the exclusion limits on $\lambda^L_{23}$ in purple ($\tau\tau$) and the $R_{D^{(*)}}$-favoured region (FlavD, which also satisfies other flavour bounds like $F_L(D^*)$, $P_\tau(D^*)$ etc.~\cite{Bhaskar:2021pml}) in yellow. Exclusion region (in magenta) come from direct searches of LQ. We find the single-coupling $R_{D^{(*)}}$ scenarios are not favoured by the LHC data. However, we see an improvement in a two-coupling $R_{D^{(*)}}$ scenario.  We show the region allowed by the LHC data in blue for the RD2A scenario for a $1.5$ TeV $U_1$ in Figure~\ref{fig:RD1allb}. There is a considerable overlap between this region and the one required by the relevant flavour observables (in yellow and green). 

In Figure \ref{fig:RK12alla}, we plot the limits on the single coupling scenario RK1A (where only $\lambda^L_{22}\neq0$). Here we see the LHC $\mu\mu$ data (in purple) is not much restrictive for the $R_{K^{(*)}}$-favoured region (FlavK). FlavK (in yellow) also consists of other flavour observables relevant to this particular scenario. However, a recast of the direct search for scalar LQ~ \cite{ATLAS:2020dsk} (magenta dotted line) effectively rules out a $1.5$ TeV LQ in the one- or two-coupling $R_{K^{(*)}}$-scenarios (Table \ref{tab:RDobs}). We show the region allowed by the LHC data for a heavier LQ in the RK2A scenario in Figure~\ref{fig:RK12allb}. 

However, a $1.5$ TeV $U_1$ can  simultaneously revolve both the anomalies. In Figure~\ref{fig:RK1500}, we show such a possibility where we have considered four new couplings to be nonzero (two relevant for the $R_{D^{(*)}}$-observable and the other two, for $R_{K^{(*)}}$).\footnote{See ref.~\cite{Mandal:2018kau} for a similar analysis with the scalar LQ $S_1$ and refs.~\cite{Chandak:2019iwj,Bhaskar:2021gsy} for the discovery prospects of LQs that couples with third generation quarks at the HL-LHC.} 

% We introduce additional coupling constants in order to make the make $\beta(U_1\to \mu b/\mu j)\lesssim0.25$. These couplings aren't new, they are allowed by the LHC data and favoured by the $R_{D^{(*)}}$ observables (See Fig. \ref{fig:RD1all} (b)). These couplings play no part in the $R_{K^{(*)}}$ observables. Thus, we obtain regions for a $1.5$ TeV LQ which can explain $R_{D^{(*)}}$ and $R_{K^{(*)}}$ observables simultaneously.

%%%%%%%%%%%%%%%%%%%%%%%%%%%%%%%%%%%%%%%%%%%%%%%%%%%%%%%%%%%%%%%%%%%%%%%%%%%%%%%%%%%%%%%%%%%%%%%%%%%%%%%%%%%%%%%%%%%%%%%%%%%%%%%%%%%%%%%%%%%%%%%%
\section{Conclusions}
%%%%%%%%%%%%%%%%%%%%%%%%%%%%%%%%%%%%%%%%%%%%%%%%%%%%%%%%%%%%%%%%%%%%%%%%
%%%%%%%%%%%%%%%%%%%%%%%%%%%%%%%%%%%%%%%%%%%%%%%%%%%%%%%%%%%%%%%%%%%%%%%%

We obtained precise bounds on the parameter space of the $U_1$ leptoquark in all possible simple $R_{D^{(*)}}$- and $R_{K^{(*)}}$- anomalies-motivated scenarios from the latest $\tau\tau$ and $\mu\mu$ resonance search data at the LHC. The current dilepton data from the LHC put stringent bounds on the $U_1$ model parameters needed to explain the $R_{D^{(*)}}$ and the $R_{K^{(*)}}$ anomalies. The obtained bounds are precise since we considered the contribution of all production modes (including the signal-background interference contribution) in the signal while recasting. In the low-mass region, the resonant modes play a significant role in obtaining the limits. We found a $1.5$ TeV LQ can explain both $R_{D^{(*)}}$ and $R_{K^{(*)}}$ anomalies simultaneously while satisfying all relevant limits.

\acknowledgments 
A.B. and S.M. acknowledge support from the Science and
Engineering Research Board (SERB), DST, India, under Grant No. ECR/2017/000517. D.D. acknowledges the DST, Government of India for the INSPIRE Faculty Fellowship (Grant No. IFA16-PH170). T.M. is supported by the intramural grant from IISER-TVM. C.N. is supported by the DST-Inspire Fellowship.

%%%%%%%%%%%%%%%%%%%%%%%%%%%%%%%%%%%%%%%%%%%%%%%%%%%%%%%%%%%%%%%%%%%%%%%%
%%%%%%%%%%%%%%%%%%%%% Bibliography %%%%%%%%%%%%%%%%%%%%%%%%%%%%%%%%%%%%%
%%%%%%%%%%%%%%%%%%%%%%%%%%%%%%%%%%%%%%%%%%%%%%%%%%%%%%%%%%%%%%%%%%%%%%%%%%%%%%%%%%%%%%%%%%%%%%%%%%%%%%%%%%%%%%%%%%%%%%%%%%%%%%%%%%%%%%%%%%%%%%%%

\bibliography{Leptoquark}{}
\bibliographystyle{JHEPCust}

\end{document}